\documentclass[%
 reprint,
 amsmath,amssymb,
 aps,
 superscriptaddress,
]{revtex4-2}

\usepackage{eqnarray}
\usepackage{xcolor}
\usepackage{graphicx}
\usepackage{dcolumn}
\usepackage{bm}

\newcommand{\micron}{\mathrm{\mu}}

\begin{document}
\title{Ultra-coherent nanomechanical resonators based on inverse design}

\author{Dennis H\o j}
	\email{denho@fysik.dtu.dk}
	\affiliation{Center for Macroscopic Quantum States (bigQ), Department of Physics, Technical University of Denmark, Fysikvej, 2800 Kgs. Lyngby, Denmark}

\author{Fengwen Wang}
	\affiliation{Department of Mechanical Engineering, Technical University of Denmark, Niels Koppels Allé, 2800 Kongens Lyngby, Denmark}

\author{Wenjun Gao}
    \affiliation{Department of Mechanical Engineering, Technical University of Denmark, Niels Koppels Allé, 2800 Kongens Lyngby, Denmark}
	\affiliation{State Key Laboratory of Disaster Reduction in Civil Engineering, Tongji University, Shanghai 200092, China}

\author{Ulrich Busk Hoff}
	\affiliation{Center for Macroscopic Quantum States (bigQ), Department of Physics, Technical University of Denmark, Fysikvej, 2800 Kgs. Lyngby, Denmark}

\author{Ole Sigmund}
	\affiliation{Department of Mechanical Engineering, Technical University of Denmark, Niels Koppels Allé, 2800 Kongens Lyngby, Denmark}

\author{Ulrik Lund Andersen}
	\email{ulrik.andersen@fysik.dtu.dk}
	\affiliation{Center for Macroscopic Quantum States (bigQ), Department of Physics, Technical University of Denmark, Fysikvej, 2800 Kgs. Lyngby, Denmark}




\begin{abstract}
Engineered micro- and nanomechanical resonators with ultra-low dissipation constitute the ideal systems for applications ranging from high-precision sensing such as magnetic resonance force microscopy, to quantum transduction between disparate quantum systems \cite{Aspelmeyer2014,Bowen2016}. Traditionally, the improvement of the resonator’s performance – often quantified by its Q$\cdot$f  product (where Q is quality factor and f is frequency) – through nanomechanical engineering such as dissipation dilution and strain engineering, has been driven by human intuition and insight\cite{Verbridge2008,Zwickl2008,Unterreithmeier2010,Schmid2011,Yu2012,Norte2016, Reinhardt2016, Tsaturyan2017,Ghadimi2018,Fedorov2020}. Such an approach is inefficient and leaves aside a plethora of unexplored mechanical designs that potentially achieve better performance. Here, we use a computer-aided inverse design approach known as topology optimization to structurally design mechanical resonators with optimal performance of the fundamental mechanical mode. Using the outcomes of this approach, we fabricate and characterize ultra-coherent nanomechanical resonators with record-high Q$\cdot$f  products, entering a quantum coherent regime where coherent oscillations are observed at room temperature. Further refinements to the model describing the mechanical system are likely to improve the Q$\cdot$f  product even more. The proposed approach – which can be also used to improve phononic crystal and coupled-mode resonators – opens up a new paradigm for designing ultra-coherent micro- and nanomechanical resonators for cutting-edge technology, enabling e.g. novel experiments in fundamental physics (e.g. search for dark matter \cite{Manley2021, Carney2021} and quantum nature of gravity \cite{Bose2017,Marletto2017}) and extreme sensing of magnetic fields \cite{Rugar2004,Poggio2010}, electric fields \cite{Lahaye2009} and mass \cite{Haney2012} with unprecedented sensitivities at room temperature. 
\end{abstract}

\maketitle

Topology optimization is a computational morphogenesis procedure widely applied in engineering to determine the best possible structural design and material distributions within a prescribed design domain to maximize a set of performance targets \cite{Bendsoee2003}. Examples include  the maximization of the structural stiffness of an object under certain design and manufacturing constraints to determine the optimal design of a full-scale aeroplane wing \cite{Aage2017} or a girder of a suspension bridge \cite{Baandrup2020}, and the maximization of light concentration to develop the optimal design of nanophotonic resonators \cite{WanChrSig18}.

The basic strategy of topology optimization is to define a design domain in which material can be distributed. Material is being added to or removed from this domain, and founded on a physical model for the system, a gradient-based computational method is used to optimize the figure-of-merit. Through iterations, material is gradually redistributed towards the optimal design for which the figure of merit is either maximized or minimized, depending on the problem to be solved.     

We use topology optimization to optimally design a nanomechanical resonator towards maximizing its Q$\cdot$f (Qf) product \cite{Gerard2017,Fu2019,Gao2020}. Previously, improving the resonator’s performance has been done through a combination of human intuition and trial-and-error based on experience and approximative analytical expression for the different dissipation mechanisms of the resonator. Such an intuition-based approach has recently led to impressive progress in increasing the Qf product of mechanical resonators by using a combination of dissipation dilution \cite{Schmid2011}, soft-clamping \cite{Tsaturyan2017}, thin-clamping \cite{Bereyhi2019} and strain engineering \cite{Ghadimi2018}. Despite these recent successes, the approach inevitably leaves out many, possibly counter-intuitive, designs that might exhibit superior behavior. Topology optimization counteracts this problem as it directly develops the optimized structure under given initial design constraints and loss models with no geometrical pre-assumptions. 

\begin{figure}[t]
\centering
\includegraphics{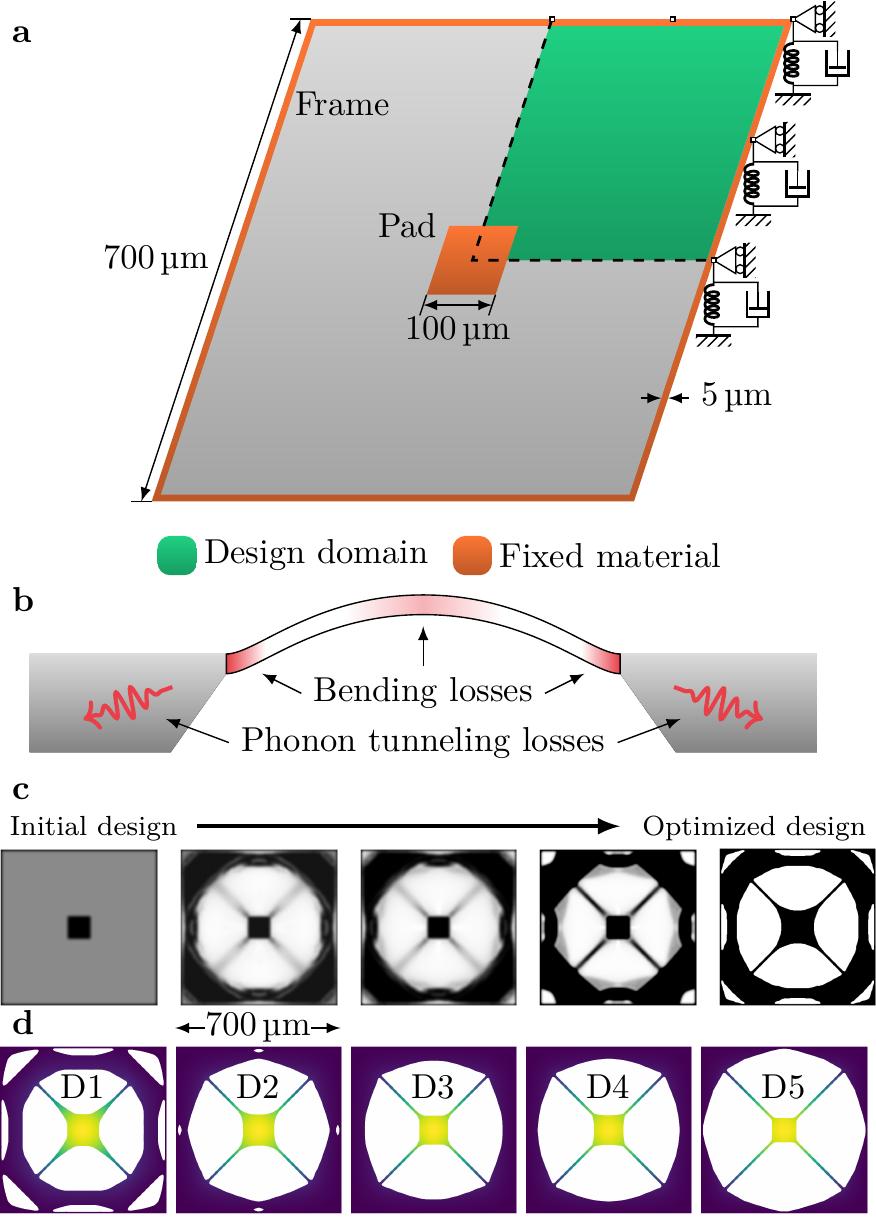}
\caption{\textbf{a.} Illustration of the model used in topology optimization. Note the springs illustrate a continuous distribution of springs. \textbf{b.} Illustration of the two damping mechanisms: Intrinsic losses in the form of bending and phonon tunneling losses. \textbf{c.} Illustration of the optimization procedure of resonator D1 with snapshots of the design evolution. The degree of transparency indicates the material density. \textbf{d.} Overview of topology optimized trampolines and the mode shape of their respective fundamental mode.}
\label{fig:TO_overview}
\end{figure}

Aiming at maximizing the Qf product of the fundamental mode of a nanomechanical resonator suitable for opto-mechanical experiments, we consider the pre-constrained structure illustrated in Fig. \ref{fig:TO_overview}a. It comprises an area of $700\times700 \ \mathrm{\mu m^2}$ with a single pad of size $100\times 100 \ \mathrm{\mu m^2}$ (that allows for the interaction with light via radiation pressure force) and a narrow frame of $5 \ \mathrm{\mu m}$ to ease fabrication. The remaining space is free to evolve through topology optimization. Furthermore, we assume that the resonator is made of pre-stressed silicon nitride with a thickness of $50$ nm. The pre-stressed resonator is numerically discretized using finite (quadrilateral shell) element  method, see details in the method section.  

Two damping mechanisms associated with intrinsic losses and phonon tunneling losses have been included in the model (Fig. \ref{fig:TO_overview}b). The intrinsic losses are modeled as bending losses in the form of hysteretic damping, i.e. using a lossy Young’s modulus. The phonon tunneling loss (PTL) associated with radiation of the phonons into the substrate was modeled 
by coupling the boundary out-of-plane displacements to continuously distributed lossy springs (illustrated in Fig. \ref{fig:TO_overview}a) 
while keeping all the other degrees of freedom fixed. We calibrate the resonator models against previous measurements and from this consider five different cases associated with different ratios between intrinsic and photon tunneling loss, denoted D1-D5. For D1 and D5 the system is purely limited by intrinsic loss and by phonon tunneling loss, respectively, while for D2-D4 the ratio is gradually changed. The exact ratios can be found in the method section. To illustrate the iterative procedure of the topology optimization, in Fig. \ref{fig:TO_overview}c we show the evolution of the design of resonator D1. The final topology optimized designs for all five cases are illustrated in Fig. \ref{fig:TO_overview}d. The images have been slightly filtered in post-processing with the aim of removing buckling-prone features and smoothing sharp features to prevent high tensile stresses at the boundaries. 

We note, interestingly, that the optimised design of D1 is similar to the membrane design suggested and experimentally tested in Ref. \cite{Beccari2021} but using a completely different approach. They arrive at this geometry based on considerations on soft-clamping using a hierarchical design concept \cite{Fedorov2020}.

\begin{figure}[t]
\centering
\includegraphics{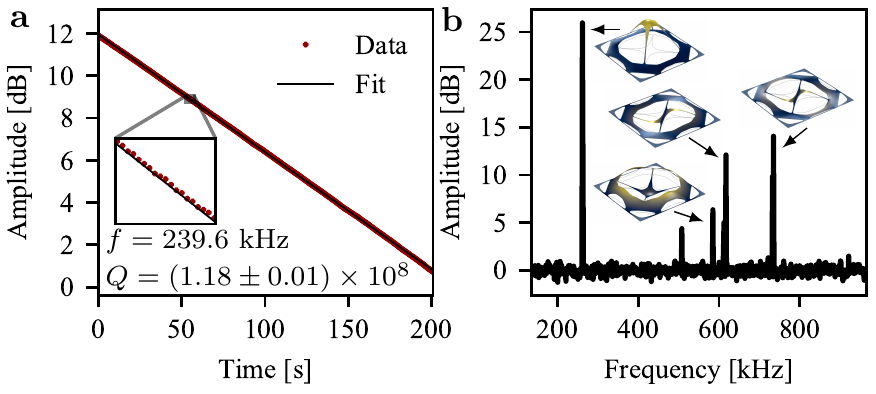}
\caption{\textbf{a.} Mechanical ringdown measurement of the best measured sample corresponding to design D4. \textbf{b.} Spectrum of a D1 sample.}
\label{fig:TOT_ringdown}
\end{figure}

\begin{figure*}[t]
\centering
\includegraphics{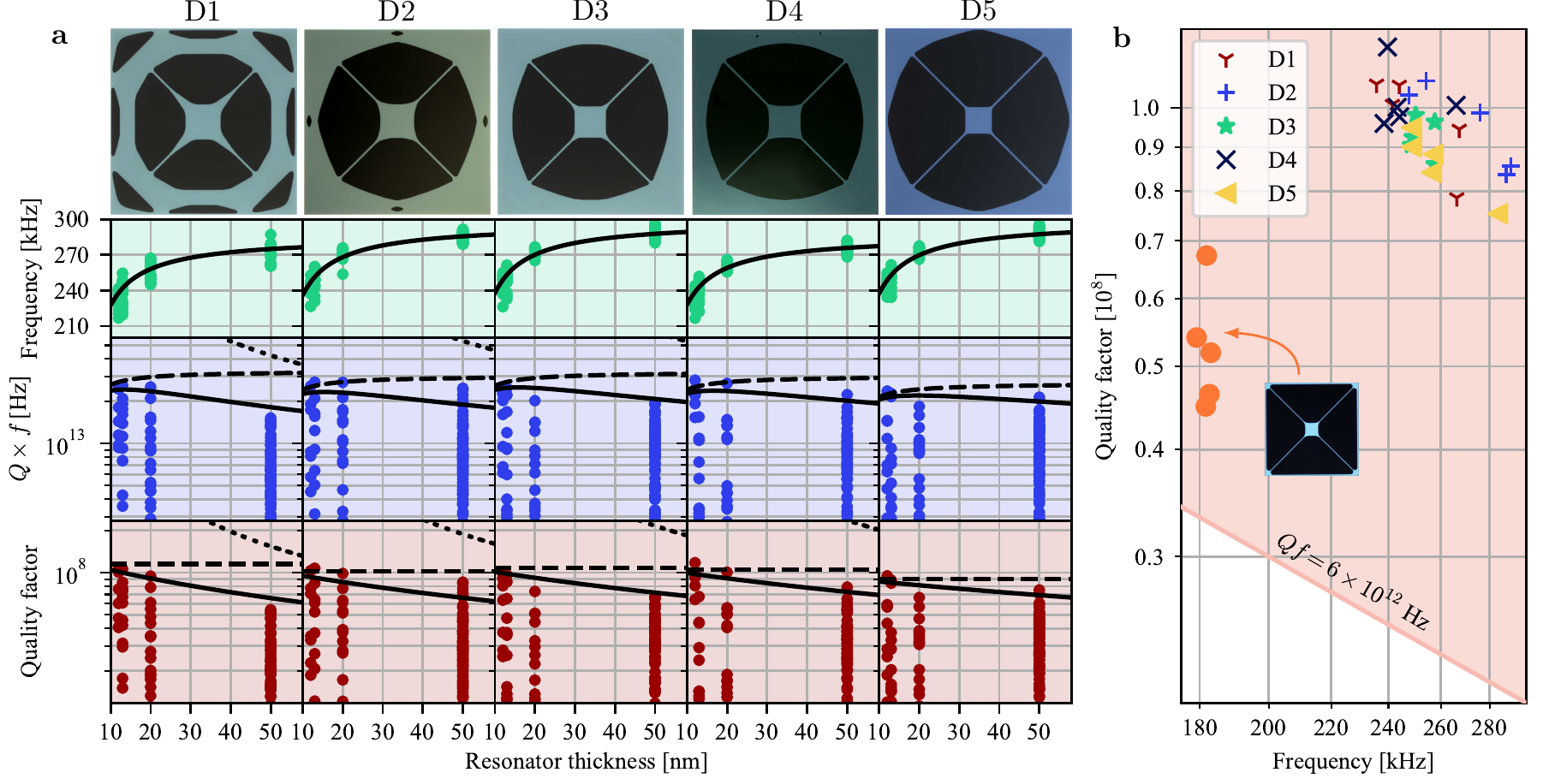}
\caption{\textbf{a.} Overview of the measured frequencies and quality factors across all designs and thicknesses together with selected microscope images. Solid lines correspond to theory fitted to the measured frequencies and best attained quality factors. Dotted and dashed lines are associated with the phonon tunneling and intrinsic loss contributions, respectively. For some designs, the theory curves for the phonon tunnelling loss is not visible on the plots. \textbf{b.} Quality factor plotted against frequency for the five best samples for each design. The shaded area marks the parameter regime in which the resonator may undergo quantum coherent oscillations at room temperature. The inset is the nominal trampoline design fabricated in this work and used as a reference \cite{Norte2016}.}
\label{fig:TOT_exp}
\end{figure*}

The post-processed designs were patterned on high-stress ($\sigma_0 \leq 1.2$ GPa) silicon nitride with a thickness of 12-50 nm grown by low pressure chemical vapor deposition on a silicon wafer. We release the resonators by back-etching the silicon substrate in a window of $1.4\times 1.4\ \mathrm{mm^2}$ (see methods). The fabricated structures are shown in Fig. \ref{fig:TOT_exp}a. To measure the mechanical frequency and quality of the fundamental mode, ring-down measurements were carried out in vacuum (pressure less than $\mathrm{10^{-7}}$ mbar) at room temperature using high-sensitivity fiber-based homodyne detection (see methods). An example of a ring-down measurement of a fundamental mode of frequency 240 kHz exhibiting an amplitude ring-down time of about 160s is illustrated in Fig. \ref{fig:TOT_ringdown}a. This corresponds to a Q factor of $1.18\pm 0.01\times 10^8$ and a Qf product of $2.83\times 10^{13}$. We also present an example of a thermal noise spectrum including some higher-order modes in Fig. \ref{fig:TOT_ringdown}b. 

We performed ring-down measurements of the fundamental mode of 967 devices that include all the topologically optimized resonators, D1-D5, as well as the conventional non-optimized trampoline resonator \cite{Reinhardt2016,Norte2016} which is used as reference structure. A collection of some of our measurements on frequency, quality factor and Qf product is presented in Fig. \ref{fig:TOT_exp}. 
It is clear from these measurements that the topologically optimized resonators are superior to the reference trampolines and that they are all deeply into the regime where the resonator is able to undergo coherent oscillations (corresponding to $\mathrm{Qf>6\times10^{12}}$ Hz) as required for quantum coherent experiments \cite{Aspelmeyer2014}. 

To understand the limitations of their performance, we fit our best attained result to a theory for the intrinsic and phonon tunneling losses. As the intrinsic loss $\Delta W$ is mainly dominated by the clamping losses near the boundaries, we directly estimate these losses from the expression 
\begin{eqnarray}
\Delta W=\int\frac{\pi\phi}{12}\frac{Eh^3 }{1-\nu^2}\left(\frac{\partial^2u}{\partial x^2}+\frac{\partial^2u}{\partial y^2}\right)^2 \, \mathrm{d}x\mathrm{d}y
\end{eqnarray}
where $h$, $E$, $\nu$ are the thickness, Young's modulus and Poisson’s ratio of the resonator material \cite{Tsaturyan2017}. $u(x,y)$ is the mode shape and the loss angle is modelled as $\phi = 1/(h\beta)$ where $\beta$ is related to the intrinsic damping at the surface. The mode profiles of all designs are simulated using the COMSOL Multiphysics package with the results shown in Fig. \ref{fig:TO_overview}c. Phonon tunneling losses are simulated by coupling the resonator to the substrate using a spring with the complex spring constant defined as
\begin{eqnarray}
k_\textrm{PTL}=-m''S\omega\left(\frac{\sqrt{6}}{3n_s}+\mathrm{i}\frac{\omega}{Q_s}\right)
\label{eq:k_PTL}
\end{eqnarray}
where $S$, $m''$, $n_s$ and $Q_s$ are the substrate's area, mass per unit area, modal density and intrinsic quality factor (see methods). These two loss contributions (intrinisic and phonon tunnelling losses) are then adjusted to match the best experimental data using the loss factors $\beta$ and $Q_s$ as fitting parameters. We find $\beta=(2.93\pm 0.19)\times 10^{11} \ \mathrm{m^{-1}}$ and $Q_s=(1.27\pm 0.31)\times 10^5$, and the resulting theory curves for all designs are shown in Fig. \ref{fig:TOT_exp}a where dotted (dashed) lines correspond to phonon tunneling (intrinsic) losses while the total contribution is represented by solid lines. It is clear that the best performing resonators of all five designs are mainly dominated by intrinsic losses. However, for some resonators we observe markedly lower performance which we attribute to a near-resonant coupling to the substrate modes, consequently leading to significantly higher phonon tunnelling losses which eventually become the dominating loss factor. This randomized coupling to the substrate modes can be circumvented by inserting a damping shield encapsulating the resonator \cite{Borrielli2016}.

\begin{figure}[t]
\centering
\includegraphics{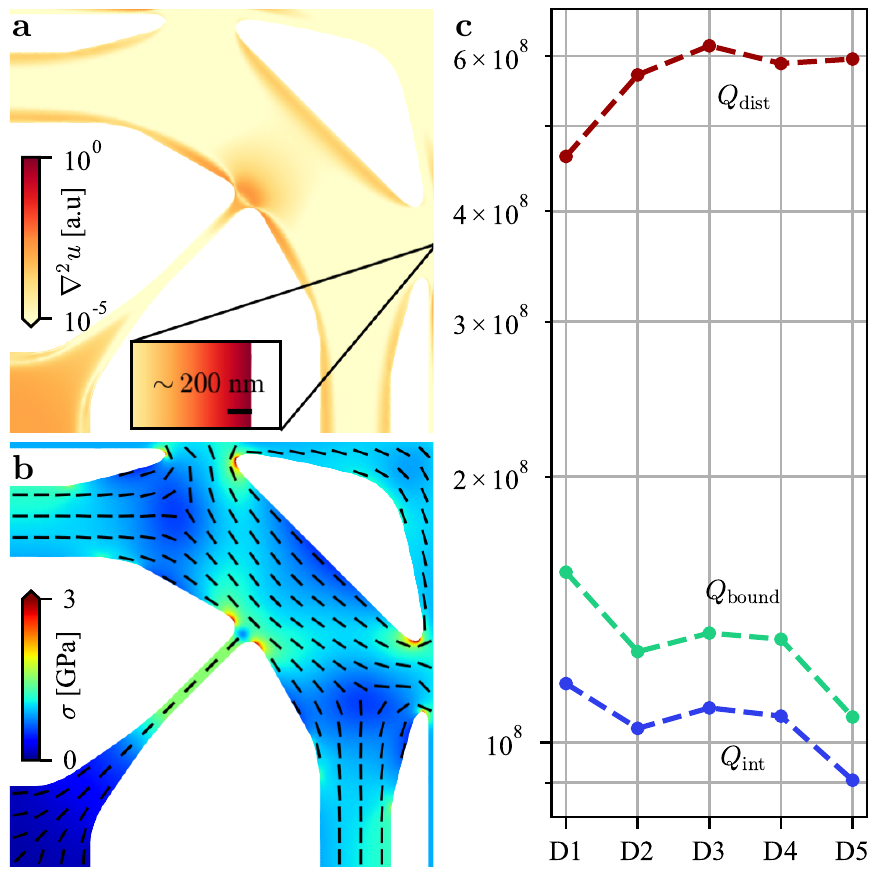}
\caption{\textbf{a.} Bending loss distribution of D1 on a logarithmic scale. The inset highlights the high bending losses at the boundary. \textbf{b.} Static von Mises distribution. The bars indicate the direction of the first principal stress component. \textbf{c.} Numerically predicted intrinsic quality factor $Q_\textrm{int}$ partitioned into boundary ($Q_\textrm{bound}$) and distributed ($Q_\textrm{dist}$) bending losses with $Q_\textrm{int}^{-1} = Q_\textrm{bound}^{-1} + Q_\textrm{dist}^{-1}$.}
\label{fig:TO_designs}
\end{figure}

We highlight the source of intrinsic losses by plotting the bending loss distribution of design D1 in Fig. \ref{fig:TO_designs}a. First we note that there is a significant amount of bending loss near the boundaries (as highlighted by the inset) and near the intersection between the circular frame and the tethers. This dilution of the bending loss into two areas (resulting from the strong mode confinement) is likely the origin of the quality enhancement, and is similar to the effect observed in resonators based on fractal structures \cite{Beccari2021}. The observed bending loss at the central pad is due to its low stress leading to a locally reduced stiffness and consequently, sharper bending. In Fig. \ref{fig:TO_designs}b we illustrate the stress distribution from which we observe a large stress component on the circular frame perpendicular to the tether. The wavelength predicted by the stress and frequency is around 2 mm which is larger than the dimensions of the resonator. Therefore, it cannot exist on the circular frame resulting in mode confinement and dilution of losses. Finally, we compared the amount of boundary bending losses (localized along the outer boundary) to the estimated amount of distributed bending losses (far away from the boundary) as shown in Fig. \ref{fig:TO_designs}c. It is clear that the resonator is limited by the boundary losses. 

Micromechanical oscillators with a Qf product of more than $10^{13}$ for the fundamental mode will have a number of intriguing applications in quantum optomechanics and precision sensing. One of the main requirements in quantum optomechanics, e.g. for cooling the oscillator to the quantum ground state, interrogating macroscopic quantum superpositions and entangling different systems, is that the decoherence time exceeds the mechanical oscillation period. This translates into the requirement that $Qf>k_BT/\hbar=6 \times 10^{12}$ Hz at room temperature (where $k_B$ is Boltzman's constant, $\hbar$ is the reduced Planck's constant and the temperature is $T=300$ K) \cite{Aspelmeyer2014,Bowen2016}. While most of the resonators fulfill this requirement, our best performing device yields around 4 coherent oscillations which is the largest number ever reported for the fundamental mode of a membrane at room temperature. Our devices will also exhibit exceptional performance in force sensing measurement as for example used in magnetic resonance force microscopy of electron and nuclear spins \cite{Fischer2019}. In such measurements the sensitivity is limited by the thermal noise ($\sqrt{4mk_BT\frac{2\pi f}{Q}}$ where $m$ is the mass) which we find to be at 10 $\mathrm{aN/\sqrt{Hz}}$ for the best devices which is significantly beyond what is attainable with currently available room temperature force microscopes.

The topology optimization method, that we have here employed to maximize the Qf product of the fundamental mode of a membrane, is applicable to many other similar morphogenesis problems in engineering of high-performance micro- and nanomechanical resonators. It can for example be applied to finding the optimal structure for maximizing the dissipation dilution effect – and thus the Qf product – in phononic crystal resonators \cite{Tsaturyan2017,Ghadimi2018} where Q factors of around one billion (at low temperature) have already been achieved without topology optimization \cite{Rossi2018}. Another interesting avenue for new studies using our methodology is to optimize other application-specific parameters instead of the Qf product. An example is the optimization of the co-operativity parameter associated with the coupling of a specifically functionalized mechanical oscillator to spins \cite{Arcizet2011}, light \cite{Verhagen2012} or charges \cite{Lahaye2009} with the aim of significantly enhancing quantum transduction or sensing. Finally, it is also possible to optimize structures with addtional constraints, either structural constrains enabling certain applications or parameter constraints, e.g.  
fixing the mass to a large value with the aim of maximizing the coupling to gravity as required for example for interrogating the quantum nature of gravity. Our methodology thus has the potential to revolutionize the way nano- and micro-mechanical systems are being designed enabling radically new applications and fundamental explorations.

\begin{acknowledgments}
This work has received funding from Villum foundation (through grant no. 13300 and Villum Investigator Project InnoTop) and the Danish National Research Foundation (bigQ DNRF142).
\end{acknowledgments}


\bibliography{Ref_UltraCoherent,apssampFW,dennis}

\providecommand{\noopsort}[1]{}\providecommand{\singleletter}[1]{#1}%
\begin{thebibliography}{41}%
\makeatletter
\providecommand \@ifxundefined [1]{%
 \@ifx{#1\undefined}
}%
\providecommand \@ifnum [1]{%
 \ifnum #1\expandafter \@firstoftwo
 \else \expandafter \@secondoftwo
 \fi
}%
\providecommand \@ifx [1]{%
 \ifx #1\expandafter \@firstoftwo
 \else \expandafter \@secondoftwo
 \fi
}%
\providecommand \natexlab [1]{#1}%
\providecommand \enquote  [1]{``#1''}%
\providecommand \bibnamefont  [1]{#1}%
\providecommand \bibfnamefont [1]{#1}%
\providecommand \citenamefont [1]{#1}%
\providecommand \href@noop [0]{\@secondoftwo}%
\providecommand \href [0]{\begingroup \@sanitize@url \@href}%
\providecommand \@href[1]{\@@startlink{#1}\@@href}%
\providecommand \@@href[1]{\endgroup#1\@@endlink}%
\providecommand \@sanitize@url [0]{\catcode `\\12\catcode `\$12\catcode
  `\&12\catcode `\#12\catcode `\^12\catcode `\_12\catcode `\%12\relax}%
\providecommand \@@startlink[1]{}%
\providecommand \@@endlink[0]{}%
\providecommand \url  [0]{\begingroup\@sanitize@url \@url }%
\providecommand \@url [1]{\endgroup\@href {#1}{\urlprefix }}%
\providecommand \urlprefix  [0]{URL }%
\providecommand \Eprint [0]{\href }%
\providecommand \doibase [0]{https://doi.org/}%
\providecommand \selectlanguage [0]{\@gobble}%
\providecommand \bibinfo  [0]{\@secondoftwo}%
\providecommand \bibfield  [0]{\@secondoftwo}%
\providecommand \translation [1]{[#1]}%
\providecommand \BibitemOpen [0]{}%
\providecommand \bibitemStop [0]{}%
\providecommand \bibitemNoStop [0]{.\EOS\space}%
\providecommand \EOS [0]{\spacefactor3000\relax}%
\providecommand \BibitemShut  [1]{\csname bibitem#1\endcsname}%
\let\auto@bib@innerbib\@empty
\bibitem [{\citenamefont {Aspelmeyer}\ \emph {et~al.}(2014)\citenamefont
  {Aspelmeyer}, \citenamefont {Kippenberg},\ and\ \citenamefont
  {Marquardt}}]{Aspelmeyer2014}%
  \BibitemOpen
  \bibfield  {author} {\bibinfo {author} {\bibfnamefont {M.}~\bibnamefont
  {Aspelmeyer}}, \bibinfo {author} {\bibfnamefont {T.~J.}\ \bibnamefont
  {Kippenberg}},\ and\ \bibinfo {author} {\bibfnamefont {F.}~\bibnamefont
  {Marquardt}},\ }\bibfield  {title} {\bibinfo {title} {{Cavity
  optomechanics}},\ }\href {https://doi.org/10.1103/RevModPhys.86.1391}
  {\bibfield  {journal} {\bibinfo  {journal} {Reviews of Modern Physics}\
  }\textbf {\bibinfo {volume} {86}},\ \bibinfo {pages} {1391} (\bibinfo {year}
  {2014})},\ \Eprint {https://arxiv.org/abs/0712.1618} {arXiv:0712.1618}
  \BibitemShut {NoStop}%
\bibitem [{\citenamefont {Bowen}\ and\ \citenamefont
  {Milburn}(2016)}]{Bowen2016}%
  \BibitemOpen
  \bibfield  {author} {\bibinfo {author} {\bibfnamefont {W.~P.}\ \bibnamefont
  {Bowen}}\ and\ \bibinfo {author} {\bibfnamefont {G.~J.}\ \bibnamefont
  {Milburn}},\ }\href@noop {} {\emph {\bibinfo {title} {{Quantum
  Optomechanics}}}},\ \bibinfo {edition} {1st}\ ed.\ (\bibinfo  {publisher}
  {CRC Press},\ \bibinfo {year} {2016})\BibitemShut {NoStop}%
\bibitem [{\citenamefont {Verbridge}\ \emph {et~al.}(2008)\citenamefont
  {Verbridge}, \citenamefont {Ilic}, \citenamefont {Craighead},\ and\
  \citenamefont {Parpia}}]{Verbridge2008}%
  \BibitemOpen
  \bibfield  {author} {\bibinfo {author} {\bibfnamefont {S.~S.}\ \bibnamefont
  {Verbridge}}, \bibinfo {author} {\bibfnamefont {R.}~\bibnamefont {Ilic}},
  \bibinfo {author} {\bibfnamefont {H.~G.}\ \bibnamefont {Craighead}},\ and\
  \bibinfo {author} {\bibfnamefont {J.~M.}\ \bibnamefont {Parpia}},\ }\bibfield
   {title} {\bibinfo {title} {{Size and frequency dependent gas damping of
  nanomechanical resonators}},\ }\href {https://doi.org/10.1063/1.2952762}
  {\bibfield  {journal} {\bibinfo  {journal} {Applied Physics Letters}\
  }\textbf {\bibinfo {volume} {93}},\ \bibinfo {pages} {1} (\bibinfo {year}
  {2008})}\BibitemShut {NoStop}%
\bibitem [{\citenamefont {Zwickl}\ \emph {et~al.}(2008)\citenamefont {Zwickl},
  \citenamefont {Shanks}, \citenamefont {Jayich}, \citenamefont {Yang},
  \citenamefont {Jayich}, \citenamefont {Thompson},\ and\ \citenamefont
  {Harris}}]{Zwickl2008}%
  \BibitemOpen
  \bibfield  {author} {\bibinfo {author} {\bibfnamefont {B.~M.}\ \bibnamefont
  {Zwickl}}, \bibinfo {author} {\bibfnamefont {W.~E.}\ \bibnamefont {Shanks}},
  \bibinfo {author} {\bibfnamefont {A.~M.}\ \bibnamefont {Jayich}}, \bibinfo
  {author} {\bibfnamefont {C.}~\bibnamefont {Yang}}, \bibinfo {author}
  {\bibfnamefont {A.~C.}\ \bibnamefont {Jayich}}, \bibinfo {author}
  {\bibfnamefont {J.~D.}\ \bibnamefont {Thompson}},\ and\ \bibinfo {author}
  {\bibfnamefont {J.~G.}\ \bibnamefont {Harris}},\ }\bibfield  {title}
  {\bibinfo {title} {{High quality mechanical and optical properties of
  commercial silicon nitride membranes}},\ }\href
  {https://doi.org/10.1063/1.2884191} {\bibfield  {journal} {\bibinfo
  {journal} {Applied Physics Letters}\ }\textbf {\bibinfo {volume} {92}},\
  \bibinfo {pages} {2006} (\bibinfo {year} {2008})},\ \Eprint
  {https://arxiv.org/abs/0711.2263} {arXiv:0711.2263} \BibitemShut {NoStop}%
\bibitem [{\citenamefont {Unterreithmeier}\ \emph {et~al.}(2010)\citenamefont
  {Unterreithmeier}, \citenamefont {Faust},\ and\ \citenamefont
  {Kotthaus}}]{Unterreithmeier2010}%
  \BibitemOpen
  \bibfield  {author} {\bibinfo {author} {\bibfnamefont {Q.~P.}\ \bibnamefont
  {Unterreithmeier}}, \bibinfo {author} {\bibfnamefont {T.}~\bibnamefont
  {Faust}},\ and\ \bibinfo {author} {\bibfnamefont {J.~P.}\ \bibnamefont
  {Kotthaus}},\ }\bibfield  {title} {\bibinfo {title} {{Damping of
  nanomechanical resonators}},\ }\href
  {https://doi.org/10.1103/PhysRevLett.105.027205} {\bibfield  {journal}
  {\bibinfo  {journal} {Physical Review Letters}\ }\textbf {\bibinfo {volume}
  {105}},\ \bibinfo {pages} {1} (\bibinfo {year} {2010})},\ \Eprint
  {https://arxiv.org/abs/1003.1868} {arXiv:1003.1868} \BibitemShut {NoStop}%
\bibitem [{\citenamefont {Schmid}\ \emph {et~al.}(2011)\citenamefont {Schmid},
  \citenamefont {Jensen}, \citenamefont {Nielsen},\ and\ \citenamefont
  {Boisen}}]{Schmid2011}%
  \BibitemOpen
  \bibfield  {author} {\bibinfo {author} {\bibfnamefont {S.}~\bibnamefont
  {Schmid}}, \bibinfo {author} {\bibfnamefont {K.~D.}\ \bibnamefont {Jensen}},
  \bibinfo {author} {\bibfnamefont {K.~H.}\ \bibnamefont {Nielsen}},\ and\
  \bibinfo {author} {\bibfnamefont {A.}~\bibnamefont {Boisen}},\ }\bibfield
  {title} {\bibinfo {title} {{Damping mechanisms in high-Q micro and
  nanomechanical string resonators}},\ }\href
  {https://doi.org/10.1103/PhysRevB.84.165307} {\bibfield  {journal} {\bibinfo
  {journal} {Physical Review B - Condensed Matter and Materials Physics}\
  }\textbf {\bibinfo {volume} {84}},\ \bibinfo {pages} {1} (\bibinfo {year}
  {2011})}\BibitemShut {NoStop}%
\bibitem [{\citenamefont {Yu}\ \emph {et~al.}(2012)\citenamefont {Yu},
  \citenamefont {Purdy},\ and\ \citenamefont {Regal}}]{Yu2012}%
  \BibitemOpen
  \bibfield  {author} {\bibinfo {author} {\bibfnamefont {P.~L.}\ \bibnamefont
  {Yu}}, \bibinfo {author} {\bibfnamefont {T.~P.}\ \bibnamefont {Purdy}},\ and\
  \bibinfo {author} {\bibfnamefont {C.~A.}\ \bibnamefont {Regal}},\ }\bibfield
  {title} {\bibinfo {title} {{Control of material damping in High-Q membrane
  microresonators}},\ }\href {https://doi.org/10.1103/PhysRevLett.108.083603}
  {\bibfield  {journal} {\bibinfo  {journal} {Physical Review Letters}\
  }\textbf {\bibinfo {volume} {108}},\ \bibinfo {pages} {1} (\bibinfo {year}
  {2012})},\ \Eprint {https://arxiv.org/abs/1111.1703} {arXiv:1111.1703}
  \BibitemShut {NoStop}%
\bibitem [{\citenamefont {Norte}\ \emph {et~al.}(2016)\citenamefont {Norte},
  \citenamefont {Moura},\ and\ \citenamefont {Gr{\"{o}}blacher}}]{Norte2016}%
  \BibitemOpen
  \bibfield  {author} {\bibinfo {author} {\bibfnamefont {R.~A.}\ \bibnamefont
  {Norte}}, \bibinfo {author} {\bibfnamefont {J.~P.}\ \bibnamefont {Moura}},\
  and\ \bibinfo {author} {\bibfnamefont {S.}~\bibnamefont {Gr{\"{o}}blacher}},\
  }\bibfield  {title} {\bibinfo {title} {{Mechanical Resonators for Quantum
  Optomechanics Experiments at Room Temperature}},\ }\href
  {https://doi.org/10.1103/PhysRevLett.116.147202} {\bibfield  {journal}
  {\bibinfo  {journal} {Physical Review Letters}\ }\textbf {\bibinfo {volume}
  {116}},\ \bibinfo {pages} {1} (\bibinfo {year} {2016})},\ \Eprint
  {https://arxiv.org/abs/1511.06235} {arXiv:1511.06235} \BibitemShut {NoStop}%
\bibitem [{\citenamefont {Reinhardt}\ \emph {et~al.}(2016)\citenamefont
  {Reinhardt}, \citenamefont {M{\"{u}}ller}, \citenamefont {Bourassa},\ and\
  \citenamefont {Sankey}}]{Reinhardt2016}%
  \BibitemOpen
  \bibfield  {author} {\bibinfo {author} {\bibfnamefont {C.}~\bibnamefont
  {Reinhardt}}, \bibinfo {author} {\bibfnamefont {T.}~\bibnamefont
  {M{\"{u}}ller}}, \bibinfo {author} {\bibfnamefont {A.}~\bibnamefont
  {Bourassa}},\ and\ \bibinfo {author} {\bibfnamefont {J.~C.}\ \bibnamefont
  {Sankey}},\ }\bibfield  {title} {\bibinfo {title} {{Ultralow-noise SiN
  trampoline resonators for sensing and optomechanics}},\ }\href
  {https://doi.org/10.1103/PhysRevX.6.021001} {\bibfield  {journal} {\bibinfo
  {journal} {Physical Review X}\ }\textbf {\bibinfo {volume} {6}},\ \bibinfo
  {pages} {1} (\bibinfo {year} {2016})},\ \Eprint
  {https://arxiv.org/abs/1511.01769} {arXiv:1511.01769} \BibitemShut {NoStop}%
\bibitem [{\citenamefont {Tsaturyan}\ \emph {et~al.}(2017)\citenamefont
  {Tsaturyan}, \citenamefont {Barg}, \citenamefont {Polzik},\ and\
  \citenamefont {Schliesser}}]{Tsaturyan2017}%
  \BibitemOpen
  \bibfield  {author} {\bibinfo {author} {\bibfnamefont {Y.}~\bibnamefont
  {Tsaturyan}}, \bibinfo {author} {\bibfnamefont {A.}~\bibnamefont {Barg}},
  \bibinfo {author} {\bibfnamefont {E.~S.}\ \bibnamefont {Polzik}},\ and\
  \bibinfo {author} {\bibfnamefont {A.}~\bibnamefont {Schliesser}},\ }\bibfield
   {title} {\bibinfo {title} {{Ultracoherent nanomechanical resonators via soft
  clamping and dissipation dilution}},\ }\href
  {https://doi.org/10.1038/nnano.2017.101} {\bibfield  {journal} {\bibinfo
  {journal} {Nature Nanotechnology}\ }\textbf {\bibinfo {volume} {12}},\
  \bibinfo {pages} {776} (\bibinfo {year} {2017})},\ \Eprint
  {https://arxiv.org/abs/1608.00937} {arXiv:1608.00937} \BibitemShut {NoStop}%
\bibitem [{\citenamefont {Engelsen}\ \emph {et~al.}(2018)\citenamefont
  {Engelsen}, \citenamefont {Ghadimi}, \citenamefont {Fedorov}, \citenamefont
  {Kippenberg}, \citenamefont {Bereyhi}, \citenamefont {Schilling},\ and\
  \citenamefont {Wilson}}]{Ghadimi2018}%
  \BibitemOpen
  \bibfield  {author} {\bibinfo {author} {\bibfnamefont {N.~J.}\ \bibnamefont
  {Engelsen}}, \bibinfo {author} {\bibfnamefont {A.~H.}\ \bibnamefont
  {Ghadimi}}, \bibinfo {author} {\bibfnamefont {S.~A.}\ \bibnamefont
  {Fedorov}}, \bibinfo {author} {\bibfnamefont {T.~J.}\ \bibnamefont
  {Kippenberg}}, \bibinfo {author} {\bibfnamefont {M.~J.}\ \bibnamefont
  {Bereyhi}}, \bibinfo {author} {\bibfnamefont {R.~D.}\ \bibnamefont
  {Schilling}},\ and\ \bibinfo {author} {\bibfnamefont {D.~J.}\ \bibnamefont
  {Wilson}},\ }\bibfield  {title} {\bibinfo {title} {{Elastic Strain
  Engineering for Ultralow Mechanical Dissipation}},\ }\href
  {https://doi.org/10.1109/OMN.2018.8454645} {\bibfield  {journal} {\bibinfo
  {journal} {International Conference on Optical MEMS and Nanophotonics}\
  }\textbf {\bibinfo {volume} {360}},\ \bibinfo {pages} {764} (\bibinfo {year}
  {2018})}\BibitemShut {NoStop}%
\bibitem [{\citenamefont {Fedorov}\ \emph {et~al.}(2020)\citenamefont
  {Fedorov}, \citenamefont {Beccari}, \citenamefont {Engelsen},\ and\
  \citenamefont {Kippenberg}}]{Fedorov2020}%
  \BibitemOpen
  \bibfield  {author} {\bibinfo {author} {\bibfnamefont {S.~A.}\ \bibnamefont
  {Fedorov}}, \bibinfo {author} {\bibfnamefont {A.}~\bibnamefont {Beccari}},
  \bibinfo {author} {\bibfnamefont {N.~J.}\ \bibnamefont {Engelsen}},\ and\
  \bibinfo {author} {\bibfnamefont {T.~J.}\ \bibnamefont {Kippenberg}},\
  }\bibfield  {title} {\bibinfo {title} {{Fractal-like Mechanical Resonators
  with a Soft-Clamped Fundamental Mode}},\ }\href
  {https://doi.org/10.1103/PhysRevLett.124.025502} {\bibfield  {journal}
  {\bibinfo  {journal} {Physical Review Letters}\ }\textbf {\bibinfo {volume}
  {124}},\ \bibinfo {pages} {25502} (\bibinfo {year} {2020})},\ \Eprint
  {https://arxiv.org/abs/1912.07439} {arXiv:1912.07439} \BibitemShut {NoStop}%
\bibitem [{\citenamefont {Manley}\ \emph {et~al.}(2021)\citenamefont {Manley},
  \citenamefont {Chowdhury}, \citenamefont {Grin}, \citenamefont {Singh},\ and\
  \citenamefont {Wilson}}]{Manley2021}%
  \BibitemOpen
  \bibfield  {author} {\bibinfo {author} {\bibfnamefont {J.}~\bibnamefont
  {Manley}}, \bibinfo {author} {\bibfnamefont {M.~D.}\ \bibnamefont
  {Chowdhury}}, \bibinfo {author} {\bibfnamefont {D.}~\bibnamefont {Grin}},
  \bibinfo {author} {\bibfnamefont {S.}~\bibnamefont {Singh}},\ and\ \bibinfo
  {author} {\bibfnamefont {D.~J.}\ \bibnamefont {Wilson}},\ }\bibfield  {title}
  {\bibinfo {title} {{Searching for vector dark matter with an optomechanical
  accelerometer}},\ }\href {https://doi.org/10.1103/physrevlett.126.061301}
  {\bibfield  {journal} {\bibinfo  {journal} {Physical Review Letters}\
  }\textbf {\bibinfo {volume} {126}},\ \bibinfo {pages} {61301} (\bibinfo
  {year} {2021})},\ \Eprint {https://arxiv.org/abs/2007.04899}
  {arXiv:2007.04899} \BibitemShut {NoStop}%
\bibitem [{\citenamefont {Carney}\ \emph {et~al.}(2021)\citenamefont {Carney},
  \citenamefont {Krnjaic}, \citenamefont {Moore}, \citenamefont {Regal},
  \citenamefont {Afek}, \citenamefont {Bhave}, \citenamefont {Brubaker},
  \citenamefont {Corbitt}, \citenamefont {Cripe}, \citenamefont {Crisosto},
  \citenamefont {Geraci}, \citenamefont {Ghosh}, \citenamefont {Harris},
  \citenamefont {Hook}, \citenamefont {Kolb}, \citenamefont {Kunjummen},
  \citenamefont {Lang}, \citenamefont {Li}, \citenamefont {Lin}, \citenamefont
  {Liu}, \citenamefont {Lykken}, \citenamefont {Magrini}, \citenamefont
  {Manley}, \citenamefont {Matsumoto}, \citenamefont {Monte}, \citenamefont
  {Monteiro}, \citenamefont {Purdy}, \citenamefont {Riedel}, \citenamefont
  {Singh}, \citenamefont {Singh}, \citenamefont {Sinha}, \citenamefont
  {Taylor}, \citenamefont {Qin}, \citenamefont {Wilson},\ and\ \citenamefont
  {Zhao}}]{Carney2021}%
  \BibitemOpen
  \bibfield  {author} {\bibinfo {author} {\bibfnamefont {D.}~\bibnamefont
  {Carney}}, \bibinfo {author} {\bibfnamefont {G.}~\bibnamefont {Krnjaic}},
  \bibinfo {author} {\bibfnamefont {D.~C.}\ \bibnamefont {Moore}}, \bibinfo
  {author} {\bibfnamefont {C.~A.}\ \bibnamefont {Regal}}, \bibinfo {author}
  {\bibfnamefont {G.}~\bibnamefont {Afek}}, \bibinfo {author} {\bibfnamefont
  {S.}~\bibnamefont {Bhave}}, \bibinfo {author} {\bibfnamefont
  {B.}~\bibnamefont {Brubaker}}, \bibinfo {author} {\bibfnamefont
  {T.}~\bibnamefont {Corbitt}}, \bibinfo {author} {\bibfnamefont
  {J.}~\bibnamefont {Cripe}}, \bibinfo {author} {\bibfnamefont
  {N.}~\bibnamefont {Crisosto}}, \bibinfo {author} {\bibfnamefont
  {A.}~\bibnamefont {Geraci}}, \bibinfo {author} {\bibfnamefont
  {S.}~\bibnamefont {Ghosh}}, \bibinfo {author} {\bibfnamefont {J.~G.}\
  \bibnamefont {Harris}}, \bibinfo {author} {\bibfnamefont {A.}~\bibnamefont
  {Hook}}, \bibinfo {author} {\bibfnamefont {E.~W.}\ \bibnamefont {Kolb}},
  \bibinfo {author} {\bibfnamefont {J.}~\bibnamefont {Kunjummen}}, \bibinfo
  {author} {\bibfnamefont {R.~F.}\ \bibnamefont {Lang}}, \bibinfo {author}
  {\bibfnamefont {T.}~\bibnamefont {Li}}, \bibinfo {author} {\bibfnamefont
  {T.}~\bibnamefont {Lin}}, \bibinfo {author} {\bibfnamefont {Z.}~\bibnamefont
  {Liu}}, \bibinfo {author} {\bibfnamefont {J.}~\bibnamefont {Lykken}},
  \bibinfo {author} {\bibfnamefont {L.}~\bibnamefont {Magrini}}, \bibinfo
  {author} {\bibfnamefont {J.}~\bibnamefont {Manley}}, \bibinfo {author}
  {\bibfnamefont {N.}~\bibnamefont {Matsumoto}}, \bibinfo {author}
  {\bibfnamefont {A.}~\bibnamefont {Monte}}, \bibinfo {author} {\bibfnamefont
  {F.}~\bibnamefont {Monteiro}}, \bibinfo {author} {\bibfnamefont
  {T.}~\bibnamefont {Purdy}}, \bibinfo {author} {\bibfnamefont {C.~J.}\
  \bibnamefont {Riedel}}, \bibinfo {author} {\bibfnamefont {R.}~\bibnamefont
  {Singh}}, \bibinfo {author} {\bibfnamefont {S.}~\bibnamefont {Singh}},
  \bibinfo {author} {\bibfnamefont {K.}~\bibnamefont {Sinha}}, \bibinfo
  {author} {\bibfnamefont {J.~M.}\ \bibnamefont {Taylor}}, \bibinfo {author}
  {\bibfnamefont {J.}~\bibnamefont {Qin}}, \bibinfo {author} {\bibfnamefont
  {D.~J.}\ \bibnamefont {Wilson}},\ and\ \bibinfo {author} {\bibfnamefont
  {Y.}~\bibnamefont {Zhao}},\ }\bibfield  {title} {\bibinfo {title}
  {{Mechanical quantum sensing in the search for dark matter}},\ }\bibfield
  {journal} {\bibinfo  {journal} {Quantum Science and Technology}\ }\textbf
  {\bibinfo {volume} {6}},\ \href {https://doi.org/10.1088/2058-9565/abcfcd}
  {10.1088/2058-9565/abcfcd} (\bibinfo {year} {2021}),\ \Eprint
  {https://arxiv.org/abs/2008.06074} {arXiv:2008.06074} \BibitemShut {NoStop}%
\bibitem [{\citenamefont {Bose}\ \emph {et~al.}(2017)\citenamefont {Bose},
  \citenamefont {Mazumdar}, \citenamefont {Morley}, \citenamefont {Ulbricht},
  \citenamefont {Toro{\v{s}}}, \citenamefont {Paternostro}, \citenamefont
  {Geraci}, \citenamefont {Barker}, \citenamefont {Kim},\ and\ \citenamefont
  {Milburn}}]{Bose2017}%
  \BibitemOpen
  \bibfield  {author} {\bibinfo {author} {\bibfnamefont {S.}~\bibnamefont
  {Bose}}, \bibinfo {author} {\bibfnamefont {A.}~\bibnamefont {Mazumdar}},
  \bibinfo {author} {\bibfnamefont {G.~W.}\ \bibnamefont {Morley}}, \bibinfo
  {author} {\bibfnamefont {H.}~\bibnamefont {Ulbricht}}, \bibinfo {author}
  {\bibfnamefont {M.}~\bibnamefont {Toro{\v{s}}}}, \bibinfo {author}
  {\bibfnamefont {M.}~\bibnamefont {Paternostro}}, \bibinfo {author}
  {\bibfnamefont {A.~A.}\ \bibnamefont {Geraci}}, \bibinfo {author}
  {\bibfnamefont {P.~F.}\ \bibnamefont {Barker}}, \bibinfo {author}
  {\bibfnamefont {M.~S.}\ \bibnamefont {Kim}},\ and\ \bibinfo {author}
  {\bibfnamefont {G.}~\bibnamefont {Milburn}},\ }\bibfield  {title} {\bibinfo
  {title} {{Spin Entanglement Witness for Quantum Gravity}},\ }\href
  {https://doi.org/10.1103/PhysRevLett.119.240401} {\bibfield  {journal}
  {\bibinfo  {journal} {Physical Review Letters}\ }\textbf {\bibinfo {volume}
  {119}},\ \bibinfo {pages} {1} (\bibinfo {year} {2017})},\ \Eprint
  {https://arxiv.org/abs/1707.06050} {arXiv:1707.06050} \BibitemShut {NoStop}%
\bibitem [{\citenamefont {Marletto}\ and\ \citenamefont
  {Vedral}(2017)}]{Marletto2017}%
  \BibitemOpen
  \bibfield  {author} {\bibinfo {author} {\bibfnamefont {C.}~\bibnamefont
  {Marletto}}\ and\ \bibinfo {author} {\bibfnamefont {V.}~\bibnamefont
  {Vedral}},\ }\bibfield  {title} {\bibinfo {title} {{Gravitationally Induced
  Entanglement between Two Massive Particles is Sufficient Evidence of Quantum
  Effects in Gravity}},\ }\href
  {https://doi.org/10.1103/PhysRevLett.119.240402} {\bibfield  {journal}
  {\bibinfo  {journal} {Physical Review Letters}\ }\textbf {\bibinfo {volume}
  {119}},\ \bibinfo {pages} {1} (\bibinfo {year} {2017})},\ \Eprint
  {https://arxiv.org/abs/1707.06036} {arXiv:1707.06036} \BibitemShut {NoStop}%
\bibitem [{\citenamefont {Rugar}\ \emph {et~al.}(2004)\citenamefont {Rugar},
  \citenamefont {Budakian}, \citenamefont {Mamin},\ and\ \citenamefont
  {Chui}}]{Rugar2004}%
  \BibitemOpen
  \bibfield  {author} {\bibinfo {author} {\bibfnamefont {D.}~\bibnamefont
  {Rugar}}, \bibinfo {author} {\bibfnamefont {R.}~\bibnamefont {Budakian}},
  \bibinfo {author} {\bibfnamefont {H.~J.}\ \bibnamefont {Mamin}},\ and\
  \bibinfo {author} {\bibfnamefont {B.~W.}\ \bibnamefont {Chui}},\ }\bibfield
  {title} {\bibinfo {title} {{Single spin detection by magnetic resonance force
  microscopy}},\ }\href {file:///Articles/2004/Rugar/Nature 2004
  Rugar.pdf%5Cnpapers3://publication/doi/10.1038/nature02658} {\bibfield
  {journal} {\bibinfo  {journal} {Nature}\ }\textbf {\bibinfo {volume} {430}},\
  \bibinfo {pages} {329} (\bibinfo {year} {2004})}\BibitemShut {NoStop}%
\bibitem [{\citenamefont {Poggio}\ and\ \citenamefont
  {Degen}(2010)}]{Poggio2010}%
  \BibitemOpen
  \bibfield  {author} {\bibinfo {author} {\bibfnamefont {M.}~\bibnamefont
  {Poggio}}\ and\ \bibinfo {author} {\bibfnamefont {C.~L.}\ \bibnamefont
  {Degen}},\ }\bibfield  {title} {\bibinfo {title} {{Force-detected nuclear
  magnetic resonance: Recent advances and future challenges}},\ }\bibfield
  {journal} {\bibinfo  {journal} {Nanotechnology}\ }\textbf {\bibinfo {volume}
  {21}},\ \href {https://doi.org/10.1088/0957-4484/21/34/342001}
  {10.1088/0957-4484/21/34/342001} (\bibinfo {year} {2010}),\ \Eprint
  {https://arxiv.org/abs/1006.3736} {arXiv:1006.3736} \BibitemShut {NoStop}%
\bibitem [{\citenamefont {Lahaye}\ \emph {et~al.}(2009)\citenamefont {Lahaye},
  \citenamefont {Suh}, \citenamefont {Echternach}, \citenamefont {Schwab},\
  and\ \citenamefont {Roukes}}]{Lahaye2009}%
  \BibitemOpen
  \bibfield  {author} {\bibinfo {author} {\bibfnamefont {M.~D.}\ \bibnamefont
  {Lahaye}}, \bibinfo {author} {\bibfnamefont {J.}~\bibnamefont {Suh}},
  \bibinfo {author} {\bibfnamefont {P.~M.}\ \bibnamefont {Echternach}},
  \bibinfo {author} {\bibfnamefont {K.~C.}\ \bibnamefont {Schwab}},\ and\
  \bibinfo {author} {\bibfnamefont {M.~L.}\ \bibnamefont {Roukes}},\ }\bibfield
   {title} {\bibinfo {title} {{Nanomechanical measurements of a superconducting
  qubit}},\ }\href {https://doi.org/10.1038/nature08093} {\bibfield  {journal}
  {\bibinfo  {journal} {Nature}\ }\textbf {\bibinfo {volume} {459}},\ \bibinfo
  {pages} {960} (\bibinfo {year} {2009})}\BibitemShut {NoStop}%
\bibitem [{\citenamefont {Hanay}\ \emph {et~al.}(2012)\citenamefont {Hanay},
  \citenamefont {Kelber}, \citenamefont {Naik}, \citenamefont {Chi},
  \citenamefont {Hentz}, \citenamefont {Bullard}, \citenamefont {Colinet},
  \citenamefont {Duraffourg},\ and\ \citenamefont {Roukes}}]{Haney2012}%
  \BibitemOpen
  \bibfield  {author} {\bibinfo {author} {\bibfnamefont {M.~S.}\ \bibnamefont
  {Hanay}}, \bibinfo {author} {\bibfnamefont {S.}~\bibnamefont {Kelber}},
  \bibinfo {author} {\bibfnamefont {A.~K.}\ \bibnamefont {Naik}}, \bibinfo
  {author} {\bibfnamefont {D.}~\bibnamefont {Chi}}, \bibinfo {author}
  {\bibfnamefont {S.}~\bibnamefont {Hentz}}, \bibinfo {author} {\bibfnamefont
  {E.~C.}\ \bibnamefont {Bullard}}, \bibinfo {author} {\bibfnamefont
  {E.}~\bibnamefont {Colinet}}, \bibinfo {author} {\bibfnamefont
  {L.}~\bibnamefont {Duraffourg}},\ and\ \bibinfo {author} {\bibfnamefont
  {M.~L.}\ \bibnamefont {Roukes}},\ }\bibfield  {title} {\bibinfo {title}
  {{Single-protein nanomechanical mass spectrometry in real time}},\ }\href
  {https://doi.org/10.1038/nnano.2012.119} {\bibfield  {journal} {\bibinfo
  {journal} {Nature Nanotechnology}\ }\textbf {\bibinfo {volume} {7}},\
  \bibinfo {pages} {602} (\bibinfo {year} {2012})}\BibitemShut {NoStop}%
\bibitem [{\citenamefont {Bends{\o}e}\ and\ \citenamefont
  {Sigmund}(2003)}]{Bendsoee2003}%
  \BibitemOpen
  \bibfield  {author} {\bibinfo {author} {\bibfnamefont {M.~P.}\ \bibnamefont
  {Bends{\o}e}}\ and\ \bibinfo {author} {\bibfnamefont {O.}~\bibnamefont
  {Sigmund}},\ }\href@noop {} {\emph {\bibinfo {title} {Topology optimization:
  theory, methods and applications}}}\ (\bibinfo  {publisher} {Springer},\
  \bibinfo {address} {Berlin},\ \bibinfo {year} {2003})\ pp.\ \bibinfo {pages}
  {1--370 s--370 s}\BibitemShut {NoStop}%
\bibitem [{\citenamefont {Aage}\ \emph {et~al.}(2017)\citenamefont {Aage},
  \citenamefont {Andreassen}, \citenamefont {Lazarov},\ and\ \citenamefont
  {Sigmund}}]{Aage2017}%
  \BibitemOpen
  \bibfield  {author} {\bibinfo {author} {\bibfnamefont {N.}~\bibnamefont
  {Aage}}, \bibinfo {author} {\bibfnamefont {E.}~\bibnamefont {Andreassen}},
  \bibinfo {author} {\bibfnamefont {B.~S.}\ \bibnamefont {Lazarov}},\ and\
  \bibinfo {author} {\bibfnamefont {O.}~\bibnamefont {Sigmund}},\ }\bibfield
  {title} {\bibinfo {title} {{Giga-voxel computational morphogenesis for
  structural design}},\ }\href {https://doi.org/10.1038/nature23911} {\bibfield
   {journal} {\bibinfo  {journal} {Nature}\ }\textbf {\bibinfo {volume}
  {550}},\ \bibinfo {pages} {84} (\bibinfo {year} {2017})}\BibitemShut
  {NoStop}%
\bibitem [{\citenamefont {Baandrup}\ \emph {et~al.}(2020)\citenamefont
  {Baandrup}, \citenamefont {Sigmund}, \citenamefont {Polk},\ and\
  \citenamefont {Aage}}]{Baandrup2020}%
  \BibitemOpen
  \bibfield  {author} {\bibinfo {author} {\bibfnamefont {M.}~\bibnamefont
  {Baandrup}}, \bibinfo {author} {\bibfnamefont {O.}~\bibnamefont {Sigmund}},
  \bibinfo {author} {\bibfnamefont {H.}~\bibnamefont {Polk}},\ and\ \bibinfo
  {author} {\bibfnamefont {N.}~\bibnamefont {Aage}},\ }\bibfield  {title}
  {\bibinfo {title} {{Closing the gap towards super-long suspension bridges
  using computational morphogenesis}},\ }\href
  {https://doi.org/10.1038/s41467-020-16599-6} {\bibfield  {journal} {\bibinfo
  {journal} {Nature Communications}\ }\textbf {\bibinfo {volume} {11}},\
  \bibinfo {pages} {1} (\bibinfo {year} {2020})}\BibitemShut {NoStop}%
\bibitem [{\citenamefont {Wang}\ \emph {et~al.}(2018)\citenamefont {Wang},
  \citenamefont {Christiansen}, \citenamefont {Yu}, \citenamefont {M{\o}rk},\
  and\ \citenamefont {Sigmund}}]{WanChrSig18}%
  \BibitemOpen
  \bibfield  {author} {\bibinfo {author} {\bibfnamefont {F.}~\bibnamefont
  {Wang}}, \bibinfo {author} {\bibfnamefont {R.}~\bibnamefont {Christiansen}},
  \bibinfo {author} {\bibfnamefont {Y.}~\bibnamefont {Yu}}, \bibinfo {author}
  {\bibfnamefont {J.}~\bibnamefont {M{\o}rk}},\ and\ \bibinfo {author}
  {\bibfnamefont {O.}~\bibnamefont {Sigmund}},\ }\bibfield  {title} {\bibinfo
  {title} {Maximizing the quality factor to mode volume ratio for ultra-small
  photonic crystal cavities},\ }\href {https://doi.org/10.1063/1.5064468}
  {\bibfield  {journal} {\bibinfo  {journal} {Applied Physics Letters}\
  }\textbf {\bibinfo {volume} {113}},\ \bibinfo {pages} {241101} (\bibinfo
  {year} {2018})}\BibitemShut {NoStop}%
\bibitem [{\citenamefont {Gerrard}\ \emph {et~al.}(2017)\citenamefont
  {Gerrard}, \citenamefont {Chen}, \citenamefont {Chandorkar}, \citenamefont
  {Yu}, \citenamefont {Rodriguez}, \citenamefont {Flader}, \citenamefont
  {Shin}, \citenamefont {Meinhart}, \citenamefont {Sigmund},\ and\
  \citenamefont {Kenny}}]{Gerard2017}%
  \BibitemOpen
  \bibfield  {author} {\bibinfo {author} {\bibfnamefont {D.~D.}\ \bibnamefont
  {Gerrard}}, \bibinfo {author} {\bibfnamefont {Y.}~\bibnamefont {Chen}},
  \bibinfo {author} {\bibfnamefont {S.~A.}\ \bibnamefont {Chandorkar}},
  \bibinfo {author} {\bibfnamefont {G.}~\bibnamefont {Yu}}, \bibinfo {author}
  {\bibfnamefont {J.}~\bibnamefont {Rodriguez}}, \bibinfo {author}
  {\bibfnamefont {I.~B.}\ \bibnamefont {Flader}}, \bibinfo {author}
  {\bibfnamefont {D.~D.}\ \bibnamefont {Shin}}, \bibinfo {author}
  {\bibfnamefont {C.~D.}\ \bibnamefont {Meinhart}}, \bibinfo {author}
  {\bibfnamefont {O.}~\bibnamefont {Sigmund}},\ and\ \bibinfo {author}
  {\bibfnamefont {T.~W.}\ \bibnamefont {Kenny}},\ }\bibfield  {title} {\bibinfo
  {title} {{Topology optimization for reduction of thermo-elastic dissipation
  in MEMS resonators}},\ }in\ \href@noop {} {\emph {\bibinfo {booktitle} {2017
  19th International Conference on Solid-State Sensors, Actuators and
  Microsystems (TRANSDUCERS)}}}\ (\bibinfo  {publisher} {IEEE},\ \bibinfo
  {address} {Kaohsiung},\ \bibinfo {year} {2017})\ pp.\ \bibinfo {pages}
  {794--797}\BibitemShut {NoStop}%
\bibitem [{\citenamefont {Fu}\ \emph {et~al.}(2019)\citenamefont {Fu},
  \citenamefont {Li},\ and\ \citenamefont {Hu}}]{Fu2019}%
  \BibitemOpen
  \bibfield  {author} {\bibinfo {author} {\bibfnamefont {Y.}~\bibnamefont
  {Fu}}, \bibinfo {author} {\bibfnamefont {L.}~\bibnamefont {Li}},\ and\
  \bibinfo {author} {\bibfnamefont {Y.}~\bibnamefont {Hu}},\ }\bibfield
  {title} {\bibinfo {title} {{Enlarging quality factor in microbeam resonators
  by topology optimization}},\ }\href
  {https://doi.org/10.1080/01495739.2018.1489744} {\bibfield  {journal}
  {\bibinfo  {journal} {Journal of Thermal Stresses}\ }\textbf {\bibinfo
  {volume} {42}},\ \bibinfo {pages} {341} (\bibinfo {year} {2019})}\BibitemShut
  {NoStop}%
\bibitem [{\citenamefont {Gao}\ \emph {et~al.}(2020)\citenamefont {Gao},
  \citenamefont {Wang},\ and\ \citenamefont {Sigmund}}]{Gao2020}%
  \BibitemOpen
  \bibfield  {author} {\bibinfo {author} {\bibfnamefont {W.}~\bibnamefont
  {Gao}}, \bibinfo {author} {\bibfnamefont {F.}~\bibnamefont {Wang}},\ and\
  \bibinfo {author} {\bibfnamefont {O.}~\bibnamefont {Sigmund}},\ }\bibfield
  {title} {\bibinfo {title} {Systematic design of high-q prestressed micro
  membrane resonators},\ }\href@noop {} {\bibfield  {journal} {\bibinfo
  {journal} {Computer Methods in Applied Mechanics and Engineering}\ }\textbf
  {\bibinfo {volume} {361}},\ \bibinfo {pages} {112692} (\bibinfo {year}
  {2020})}\BibitemShut {NoStop}%
\bibitem [{\citenamefont {Bereyhi}\ \emph {et~al.}(2019)\citenamefont
  {Bereyhi}, \citenamefont {Beccari}, \citenamefont {Fedorov}, \citenamefont
  {Ghadimi}, \citenamefont {Schilling}, \citenamefont {Wilson}, \citenamefont
  {Engelsen},\ and\ \citenamefont {Kippenberg}}]{Bereyhi2019}%
  \BibitemOpen
  \bibfield  {author} {\bibinfo {author} {\bibfnamefont {M.~J.}\ \bibnamefont
  {Bereyhi}}, \bibinfo {author} {\bibfnamefont {A.}~\bibnamefont {Beccari}},
  \bibinfo {author} {\bibfnamefont {S.~A.}\ \bibnamefont {Fedorov}}, \bibinfo
  {author} {\bibfnamefont {A.~H.}\ \bibnamefont {Ghadimi}}, \bibinfo {author}
  {\bibfnamefont {R.}~\bibnamefont {Schilling}}, \bibinfo {author}
  {\bibfnamefont {D.~J.}\ \bibnamefont {Wilson}}, \bibinfo {author}
  {\bibfnamefont {N.~J.}\ \bibnamefont {Engelsen}},\ and\ \bibinfo {author}
  {\bibfnamefont {T.~J.}\ \bibnamefont {Kippenberg}},\ }\bibfield  {title}
  {\bibinfo {title} {{Clamp-Tapering Increases the Quality Factor of Stressed
  Nanobeams}},\ }\href {https://doi.org/10.1021/acs.nanolett.8b04942}
  {\bibfield  {journal} {\bibinfo  {journal} {Nano Letters}\ }\textbf {\bibinfo
  {volume} {19}},\ \bibinfo {pages} {2329} (\bibinfo {year}
  {2019})}\BibitemShut {NoStop}%
\bibitem [{\citenamefont {Beccari}\ \emph {et~al.}(2021)\citenamefont
  {Beccari}, \citenamefont {Bereyhi}, \citenamefont {Groth}, \citenamefont
  {Fedorov}, \citenamefont {Arabmoheghi}, \citenamefont {Engelsen},\ and\
  \citenamefont {Kippenberg}}]{Beccari2021}%
  \BibitemOpen
  \bibfield  {author} {\bibinfo {author} {\bibfnamefont {A.}~\bibnamefont
  {Beccari}}, \bibinfo {author} {\bibfnamefont {M.~J.}\ \bibnamefont
  {Bereyhi}}, \bibinfo {author} {\bibfnamefont {R.}~\bibnamefont {Groth}},
  \bibinfo {author} {\bibfnamefont {S.~A.}\ \bibnamefont {Fedorov}}, \bibinfo
  {author} {\bibfnamefont {A.}~\bibnamefont {Arabmoheghi}}, \bibinfo {author}
  {\bibfnamefont {N.~J.}\ \bibnamefont {Engelsen}},\ and\ \bibinfo {author}
  {\bibfnamefont {T.~J.}\ \bibnamefont {Kippenberg}},\ }\bibfield  {title}
  {\bibinfo {title} {{Hierarchical tensile structures with ultralow mechanical
  dissipation}},\ }\Eprint {https://arxiv.org/abs/2103.09785}
  {arXiv:2103.09785}  (\bibinfo {year} {2021})\BibitemShut {NoStop}%
\bibitem [{\citenamefont {Borrielli}\ \emph {et~al.}(2016)\citenamefont
  {Borrielli}, \citenamefont {Marconi}, \citenamefont {Marin}, \citenamefont
  {Marino}, \citenamefont {Morana}, \citenamefont {Pandraud}, \citenamefont
  {Pontin}, \citenamefont {Prodi}, \citenamefont {Sarro}, \citenamefont
  {Serra},\ and\ \citenamefont {Bonaldi}}]{Borrielli2016}%
  \BibitemOpen
  \bibfield  {author} {\bibinfo {author} {\bibfnamefont {A.}~\bibnamefont
  {Borrielli}}, \bibinfo {author} {\bibfnamefont {L.}~\bibnamefont {Marconi}},
  \bibinfo {author} {\bibfnamefont {F.}~\bibnamefont {Marin}}, \bibinfo
  {author} {\bibfnamefont {F.}~\bibnamefont {Marino}}, \bibinfo {author}
  {\bibfnamefont {B.}~\bibnamefont {Morana}}, \bibinfo {author} {\bibfnamefont
  {G.}~\bibnamefont {Pandraud}}, \bibinfo {author} {\bibfnamefont
  {A.}~\bibnamefont {Pontin}}, \bibinfo {author} {\bibfnamefont {G.~A.}\
  \bibnamefont {Prodi}}, \bibinfo {author} {\bibfnamefont {P.~M.}\ \bibnamefont
  {Sarro}}, \bibinfo {author} {\bibfnamefont {E.}~\bibnamefont {Serra}},\ and\
  \bibinfo {author} {\bibfnamefont {M.}~\bibnamefont {Bonaldi}},\ }\bibfield
  {title} {\bibinfo {title} {{Control of recoil losses in nanomechanical SiN
  membrane resonators}},\ }\href {https://doi.org/10.1103/PhysRevB.94.121403}
  {\bibfield  {journal} {\bibinfo  {journal} {Physical Review B}\ }\textbf
  {\bibinfo {volume} {94}},\ \bibinfo {pages} {3} (\bibinfo {year} {2016})},\
  \Eprint {https://arxiv.org/abs/1607.04485} {arXiv:1607.04485} \BibitemShut
  {NoStop}%
\bibitem [{\citenamefont {Fischer}\ \emph {et~al.}(2019)\citenamefont
  {Fischer}, \citenamefont {McNally}, \citenamefont {Reetz}, \citenamefont
  {Assumpcao}, \citenamefont {Knief}, \citenamefont {Lin},\ and\ \citenamefont
  {Regal}}]{Fischer2019}%
  \BibitemOpen
  \bibfield  {author} {\bibinfo {author} {\bibfnamefont {R.}~\bibnamefont
  {Fischer}}, \bibinfo {author} {\bibfnamefont {D.~P.}\ \bibnamefont
  {McNally}}, \bibinfo {author} {\bibfnamefont {C.}~\bibnamefont {Reetz}},
  \bibinfo {author} {\bibfnamefont {G.~G.}\ \bibnamefont {Assumpcao}}, \bibinfo
  {author} {\bibfnamefont {T.}~\bibnamefont {Knief}}, \bibinfo {author}
  {\bibfnamefont {Y.}~\bibnamefont {Lin}},\ and\ \bibinfo {author}
  {\bibfnamefont {C.~A.}\ \bibnamefont {Regal}},\ }\bibfield  {title} {\bibinfo
  {title} {{Spin detection with a micromechanical trampoline: Towards magnetic
  resonance microscopy harnessing cavity optomechanics}},\ }\href@noop {}
  {\bibfield  {journal} {\bibinfo  {journal} {New Journal of Physics}\ }\textbf
  {\bibinfo {volume} {21}} (\bibinfo {year} {2019})}\BibitemShut {NoStop}%
\bibitem [{\citenamefont {Rossi}\ \emph {et~al.}(2018)\citenamefont {Rossi},
  \citenamefont {Mason}, \citenamefont {Chen}, \citenamefont {Tsaturyan},\ and\
  \citenamefont {Schliesser}}]{Rossi2018}%
  \BibitemOpen
  \bibfield  {author} {\bibinfo {author} {\bibfnamefont {M.}~\bibnamefont
  {Rossi}}, \bibinfo {author} {\bibfnamefont {D.}~\bibnamefont {Mason}},
  \bibinfo {author} {\bibfnamefont {J.}~\bibnamefont {Chen}}, \bibinfo {author}
  {\bibfnamefont {Y.}~\bibnamefont {Tsaturyan}},\ and\ \bibinfo {author}
  {\bibfnamefont {A.}~\bibnamefont {Schliesser}},\ }\bibfield  {title}
  {\bibinfo {title} {{Measurement-based quantum control of mechanical
  motion}},\ }\href {http://arxiv.org/abs/1805.05087} {\bibfield  {journal}
  {\bibinfo  {journal} {Nature}\ }\textbf {\bibinfo {volume} {563}},\ \bibinfo
  {pages} {53} (\bibinfo {year} {2018})},\ \Eprint
  {https://arxiv.org/abs/1805.05087} {arXiv:1805.05087} \BibitemShut {NoStop}%
\bibitem [{\citenamefont {Arcizet}\ \emph {et~al.}(2011)\citenamefont
  {Arcizet}, \citenamefont {Jacques}, \citenamefont {Siria}, \citenamefont
  {Poncharal}, \citenamefont {Vincent},\ and\ \citenamefont
  {Seidelin}}]{Arcizet2011}%
  \BibitemOpen
  \bibfield  {author} {\bibinfo {author} {\bibfnamefont {O.}~\bibnamefont
  {Arcizet}}, \bibinfo {author} {\bibfnamefont {V.}~\bibnamefont {Jacques}},
  \bibinfo {author} {\bibfnamefont {A.}~\bibnamefont {Siria}}, \bibinfo
  {author} {\bibfnamefont {P.}~\bibnamefont {Poncharal}}, \bibinfo {author}
  {\bibfnamefont {P.}~\bibnamefont {Vincent}},\ and\ \bibinfo {author}
  {\bibfnamefont {S.}~\bibnamefont {Seidelin}},\ }\bibfield  {title} {\bibinfo
  {title} {{A single nitrogen-vacancy defect coupled to a nanomechanical
  oscillator}},\ }\href {https://doi.org/10.1038/nphys2070} {\bibfield
  {journal} {\bibinfo  {journal} {Nature Physics}\ }\textbf {\bibinfo {volume}
  {7}},\ \bibinfo {pages} {879} (\bibinfo {year} {2011})}\BibitemShut {NoStop}%
\bibitem [{\citenamefont {Verhagen}\ \emph {et~al.}(2012)\citenamefont
  {Verhagen}, \citenamefont {Del{\'{e}}glise}, \citenamefont {Weis},
  \citenamefont {Schliesser},\ and\ \citenamefont {Kippenberg}}]{Verhagen2012}%
  \BibitemOpen
  \bibfield  {author} {\bibinfo {author} {\bibfnamefont {E.}~\bibnamefont
  {Verhagen}}, \bibinfo {author} {\bibfnamefont {S.}~\bibnamefont
  {Del{\'{e}}glise}}, \bibinfo {author} {\bibfnamefont {S.}~\bibnamefont
  {Weis}}, \bibinfo {author} {\bibfnamefont {A.}~\bibnamefont {Schliesser}},\
  and\ \bibinfo {author} {\bibfnamefont {T.~J.}\ \bibnamefont {Kippenberg}},\
  }\bibfield  {title} {\bibinfo {title} {{Quantum-coherent coupling of a
  mechanical oscillator to an optical cavity mode}},\ }\href
  {https://doi.org/10.1038/nature10787} {\bibfield  {journal} {\bibinfo
  {journal} {Nature}\ }\textbf {\bibinfo {volume} {482}},\ \bibinfo {pages}
  {63} (\bibinfo {year} {2012})},\ \Eprint {https://arxiv.org/abs/1107.3761}
  {arXiv:1107.3761} \BibitemShut {NoStop}%
\bibitem [{\citenamefont {Dvorkin}\ and\ \citenamefont
  {Bathe}(1984)}]{Dvorkin1984}%
  \BibitemOpen
  \bibfield  {author} {\bibinfo {author} {\bibfnamefont {E.~N.}\ \bibnamefont
  {Dvorkin}}\ and\ \bibinfo {author} {\bibfnamefont {K.}~\bibnamefont
  {Bathe}},\ }\bibfield  {title} {\bibinfo {title} {A continuum mechanics based
  four‐node shell element for general non‐linear analysis},\ }\href@noop {}
  {\bibfield  {journal} {\bibinfo  {journal} {Engineering computations}\ }
  (\bibinfo {year} {1984})}\BibitemShut {NoStop}%
\bibitem [{\citenamefont {Bourdin}(2001)}]{Bourdin2001}%
  \BibitemOpen
  \bibfield  {author} {\bibinfo {author} {\bibfnamefont {B.}~\bibnamefont
  {Bourdin}},\ }\bibfield  {title} {\bibinfo {title} {Filters in topology
  optimization},\ }\href@noop {} {\bibfield  {journal} {\bibinfo  {journal}
  {International Journal for Numerical Methods in Engineering}\ }\textbf
  {\bibinfo {volume} {50}},\ \bibinfo {pages} {2143} (\bibinfo {year}
  {2001})}\BibitemShut {NoStop}%
\bibitem [{\citenamefont {Wang}\ \emph {et~al.}(2011)\citenamefont {Wang},
  \citenamefont {Lazarov},\ and\ \citenamefont {Sigmund}}]{Wang2011}%
  \BibitemOpen
  \bibfield  {author} {\bibinfo {author} {\bibfnamefont {F.}~\bibnamefont
  {Wang}}, \bibinfo {author} {\bibfnamefont {B.~S.}\ \bibnamefont {Lazarov}},\
  and\ \bibinfo {author} {\bibfnamefont {O.}~\bibnamefont {Sigmund}},\
  }\bibfield  {title} {\bibinfo {title} {On projection methods, convergence and
  robust formulations in topology optimization},\ }\href@noop {} {\bibfield
  {journal} {\bibinfo  {journal} {Structural and Multidisciplinary
  Optimization}\ }\textbf {\bibinfo {volume} {43}},\ \bibinfo {pages} {767}
  (\bibinfo {year} {2011})}\BibitemShut {NoStop}%
\bibitem [{\citenamefont {Stolpe}\ and\ \citenamefont
  {Svanberg}(2001)}]{Stolpe2001}%
  \BibitemOpen
  \bibfield  {author} {\bibinfo {author} {\bibfnamefont {M.}~\bibnamefont
  {Stolpe}}\ and\ \bibinfo {author} {\bibfnamefont {K.}~\bibnamefont
  {Svanberg}},\ }\bibfield  {title} {\bibinfo {title} {An alternative
  interpolation scheme for minimum compliance topology optimization},\
  }\href@noop {} {\bibfield  {journal} {\bibinfo  {journal} {Structural and
  Multidisciplinary Optimization}\ }\textbf {\bibinfo {volume} {22}},\ \bibinfo
  {pages} {116} (\bibinfo {year} {2001})}\BibitemShut {NoStop}%
\bibitem [{\citenamefont {Svanberg}(1987)}]{Svanberg1987}%
  \BibitemOpen
  \bibfield  {author} {\bibinfo {author} {\bibfnamefont {K.}~\bibnamefont
  {Svanberg}},\ }\bibfield  {title} {\bibinfo {title} {The method of moving
  asymptotes--a new method for structural optimization},\ }\href@noop {}
  {\bibfield  {journal} {\bibinfo  {journal} {International journal for
  numerical methods in engineering}\ }\textbf {\bibinfo {volume} {24}},\
  \bibinfo {pages} {359} (\bibinfo {year} {1987})}\BibitemShut {NoStop}%
\bibitem [{\citenamefont {Cremer}\ \emph {et~al.}(2005)\citenamefont {Cremer},
  \citenamefont {Heckl},\ and\ \citenamefont {Petersson}}]{Cremer2005}%
  \BibitemOpen
  \bibfield  {author} {\bibinfo {author} {\bibfnamefont {L.}~\bibnamefont
  {Cremer}}, \bibinfo {author} {\bibfnamefont {M.}~\bibnamefont {Heckl}},\ and\
  \bibinfo {author} {\bibfnamefont {B.~A.}\ \bibnamefont {Petersson}},\ }\href
  {https://doi.org/10.1007/b137728} {\emph {\bibinfo {title} {{Structure-borne
  sound: Structural vibrations and sound radiation at audio frequencies}}}}\
  (\bibinfo  {publisher} {Sprigne- Verlag},\ \bibinfo {address} {Heidelberg},\
  \bibinfo {year} {2005})\ pp.\ \bibinfo {pages} {1--607}\BibitemShut {NoStop}%
\bibitem [{\citenamefont {Raider}\ \emph {et~al.}(1976)\citenamefont {Raider},
  \citenamefont {Flitsch}, \citenamefont {Aboaf},\ and\ \citenamefont
  {Pliskin}}]{Raider1976}%
  \BibitemOpen
  \bibfield  {author} {\bibinfo {author} {\bibfnamefont {S.~I.}\ \bibnamefont
  {Raider}}, \bibinfo {author} {\bibfnamefont {R.}~\bibnamefont {Flitsch}},
  \bibinfo {author} {\bibfnamefont {J.~A.}\ \bibnamefont {Aboaf}},\ and\
  \bibinfo {author} {\bibfnamefont {W.~A.}\ \bibnamefont {Pliskin}},\
  }\bibfield  {title} {\bibinfo {title} {{Surface Oxidation of Silicon Nitride
  Films}},\ }\href {https://doi.org/10.1149/1.2132877} {\bibfield  {journal}
  {\bibinfo  {journal} {Journal of The Electrochemical Society}\ }\textbf
  {\bibinfo {volume} {123}},\ \bibinfo {pages} {560} (\bibinfo {year}
  {1976})}\BibitemShut {NoStop}%
\end{thebibliography}%

\section{Methods}
\subsection{Topology optimization implementation}
We employed a density-based topology optimization approach~\cite{Bendsoee2003}  to design ultrahigh coherent resonators.  The basic methodology and the detailed optimization formulation  are described in the following.

Prestressed membrane resonators are simulated using finite element methods with the 4-node MITC (Mixed Interpolation of Tensorial Components) quadrilateral shell element~\cite{Dvorkin1984}. The mechanical dynamic problem is solved in two steps: 1) Establish static equilibrium of a prestressed membrane resonator under prescribed stress;  2) Identify resonating modes  using linear eigenvalue analysis. The FE equations are stated in discrete form as,
\begin{align}
\boldsymbol{K}_0\boldsymbol{U}_0& =\boldsymbol{F}_0 \label{equ:state}\\
\left( \boldsymbol{K}_0+\boldsymbol{K}_{\sigma}\left( \boldsymbol{U}_0\right) + i\boldsymbol{C} -\omega^2_j \boldsymbol{M}\right) \boldsymbol{\phi}_j&=\boldsymbol{0}. \label{equ:Eigs}
\end{align}
Here $\boldsymbol{F}_0$ is the equivalent force vector resulting from a prestress $\sigma_0$, $\boldsymbol{K}_0$ represents the linear stiffness matrix and $\boldsymbol{K}_{\sigma}\left( \boldsymbol{U}_0\right) $  represents the initial stress stiffness matrix that depends on the displacement $\boldsymbol{U}_0$ of the prestress problem in Eq.~\eqref{equ:state}.  $\boldsymbol{C}$ and $\boldsymbol{M}$  denote damping and mass  matrixes,  $\omega_j$ and $\boldsymbol{\phi}_j$ are the angular frequency and modal profile of the $j$-th resonating mode and $i=\sqrt{-1}$  is the imaginary unit.

The damping matrix, $\boldsymbol{C}$, covers intrinsic and phonon tunneling losses.  The intrinsic losses are considered via a relaxation mechanism described by a complex-valued Young’s modulus $\tilde{E} = \left(1 + i\eta_s\right)E$. The phonon tunneling losses are modeled using damped springs distributed along the boundary with a total stiffness of $\bar{k}_b=\left( 1+i \eta_b\right)k_b$ and $k_b=8.315\times 10^7\ \rm{kN/m^2}$. The detailed calculation formulations of quantities in Eqs~\eqref{equ:state}~ and~\eqref{equ:Eigs} can be found in~\cite{Gao2020}.   The quality factor and frequency of the $j$-th resonating mode are calculated by
\begin{align}
Q_j=\frac{\Re{\left( \omega_j\right) }}{2\Im{\left( \omega_j\right) }},\quad f_j=\frac{\Re{\left( \omega_j\right) }}{2\pi}.
\end{align}

In the density-based topology optimization approach, an element-wise  design variable, $x_e\in \left[ 0, \ 1\right]$, is introduced to indicate the material occupation in element $e$.  To avoid checkerboard pattern  and mesh dependence~\cite{Bendsoee2003} and enhance design discreteness, the design variables are first filtered using a density filter~\cite{Bourdin2001}  and then smoothly projected using a  hyperbolic tangent threshold function~\cite{Wang2011}, given as
\begin{align} \label{Eq:2DProj}
\tilde{x}_{e}=&\frac{\sum\limits_{k\in N_{e}} {w_{e}}(\bm y_{k})v_{k}x_{k}}
{\sum\limits_{k\in N_{e}}w_{e}(\bm y_{k})v_{k}}\\
\bar{x}_e=&\frac{\tanh{\left(\beta_1\eta\right)}+\tanh{\left(\beta_1\left(\tilde{x}_e-\eta\right)\right)}}{\tanh{\left(\beta_1\eta\right)}+\tanh{\left(\beta_1\left(1-\eta\right)\right)}}. 
\end{align}
Here,  $\tilde{x}_e$ is the filtered design variable, $\bm y_{k}$ are the center coordinates of element $k$. $v_{k}$ and $x_{k}$ are the corresponding volume and design variable  of element $k$, respectively.~$N_{e}$ is the neighborhood of element $e$ within a certain filter radius specified by ~$N_{e}=\left\{k\middle|\,\|\bm x_{k}-\bm y_{e}\|\le r\right\}$, and~$w_{e}(\bm y_{k})=r-\|\bm y_{k}-\bm y_{e}\|$. $\bar{x}_e$ is the projected design variable of element, $e$.  When $\beta_1$ is large, $\bar{x}_e\approx 1$ if $\tilde{x}_e>\eta$ representing $\mathrm{Si_3N_4}$,  and $\bar{x}_e\approx0$ if $\tilde{x}_e<\eta$ indicating void.  The projection   suppresses  gray element density regions induced by the density filter when  $\beta_1$ is sufficiently large and ensures black-white designs when the optimization converges.  Moreover, it mimics the manufacturing process and the manufacturing errors can be taken into accounts in the optimization  by choosing different thresholds, $\eta$, as discussed later.  


The Young’s modulus   of element $e$  is directly
related to the projected  design variable using  the Rational Approximation of Material Properties (RAMP)~\cite{Stolpe2001}  and the mass density is linearly interpolated as
\begin{align} \label{Eq:2DProj_a}
{E}_e=&\frac{\bar{x}_e}{1+q\left(1-\bar{x}_e \right) }( {E}- {E}_0)+ {E}_0, \quad q=3\\ 
\rho_e=&\bar{x}_e \left( \rho- \rho_0\right)+\rho_0.
\end{align}
Spurious modes caused by inappropriate stiffness-to-mass ratios in low-density regions are suppressed by setting    ${E}_0=10^{-6}E$ and $\rho_0=10^{-7} \rho $ to represent void in this study. Wrinkling-like instabilities in low-density regions are alleviated using a displacement interpolation with detailed formulations presented in~\cite{Gao2020}. 

To enhance the design robustness  with respect to manufacturing errors and impose a minimal length scale in the nominal design, a three-case robust formulation is employed~\cite{Wang2011}. Three design realizations are generated to mimic an eroded, normal and dilated manufacturing processes. The optimization problem for designing ultrahigh coherent resonators is formulated  to maximize  the Qf product of  the fundamental mode  for the worst case of the three design realizations, subjected to   frequency constraints and a volume fraction constraint,  given as
\begin{align}  
\max\limits_{\boldsymbol{x}}    &  \qquad  \min\limits_{\eta} \qquad  \ln\left(Q_1\left(\boldsymbol{x},\eta\right)f_1\left(\boldsymbol{x},\eta\right) \right)   \nonumber \\
s.t. & \qquad f_1\left(\boldsymbol{x},\eta\right)>f^*   \nonumber\\
&  \qquad \frac{\boldsymbol{v}^T\bar{\boldsymbol{x}} \left(\boldsymbol{x},\eta_d\right)}{\sum\limits_e v_e}   \leq v^*  \nonumber \\
&  \qquad  \boldsymbol{0} \leq \boldsymbol{x} \leq \boldsymbol{1} \nonumber \\
&   \qquad  \eta \in\left\lbrace \eta_e,  \eta_i, \eta_d \right\rbrace \nonumber 
\label{eq:opt} 
\end{align}
The three design realizations are generated using $\eta\in \left\lbrace 0.55, \ 0.5, \ 0.45 \right\rbrace $ with a filter radius of $r= 15 \ \mathrm{\mu m}$.  This corresponds to a minimal feature size of 6.7 $\mathrm{\mu m}$ in both solid and void regions of the nominal design. The prescribed frequency lower bound and volume fraction upper bound are $f^*=240$   kHz and $v^*=0.5 $.

Gradients of the objective and constraint functions are calculated using the adjoint sensitivity analysis and the chain rules~\cite{Bourdin2001,Gao2020,Wang2011}. The design variables are iteratively updated using the deterministic  mathematical programming approach,  Method of Moving Asymptotes (MMA) \cite{Svanberg1987} based on the gradients of the objective and constraints. $\beta_1$ is updated until the convergence criterion is satisfied by $\beta_1^{(n+1)}=1.1\beta_1^{(n)}$ reaching a maximum value of 120.

The loss parameters used in the five design cases, (D1, D2, D3, D4, D5), are $\eta_s$=(2.500; 1.790; 1.120; 0.515; 0.000)10$^{-4}$ and $\eta_b$=(0.000; 0.095; 0.190; 0.285; 0.380) calibrated against the reference trampoline design.
\
 

\subsection{Fabrication}
We deposit stoichiometric silicon nitride onto a 100 mm single-crystal silicon wafer of 500 $\micron$m thickness using low-pressure chemical vapour deposition. This is followed by spincoating photoresist onto the wafer and transfer of the different resonator designs using UV-lithography. The photoresist is developed and the silicon nitride is etched in these regions by means of reactive ion etching. Residual photoresist is removed using oxide plasma, and finally, the trampolines are released in potassium hydroxide at $\mathrm{80^\circ C}$ followed by cleaning in hydrochloric acid and sulfiric acid mixed with ammonium persulfate. 
 
\subsection{Characterization}
Measurements of the frequency and quality factor of the resonators are performed using optical interferometry driven by a laser with a wavelength of 1550 nm. The laser beam is reflected off the vibrating membrane (located inside a vacuum chamber at low pressure $<\mathrm{10^{-7} \ mBar}$), and the resulting phase shift is detected with high-sensitivity using a phase-locked homodyne detector and recorded with a spectrum analyzer. Excitation of the mechanical oscillator is done by modulating the intensity of the laser at the resonance frequency. Once excited, the modulation is switched off and the amplitude decay is subsequently measured. 


\subsection{Phonon tunneling loss model}
The implemented phonon tunneling loss model used in evaluating the designs post topology optimization, is derived by treating the out-of-plane forces induced by the resonator onto the substrate as a point source on a large but finite silicon substrate. This assumption is valid as long as the characteristic size of the membrane as well as the substrate thickness are much smaller than the wavelength of the radiating wave in the substrate. This is the case for the treated designs when considering their fundamental mode. Furthermore, at these low frequencies, Kirchoff-Love plate theory can be used to model the vibrations induced into the substrate.

In the point-source assumption the amplitude of vibrations $u_s$ at an excited point $(x,y)$ on the finite substrate is related to the excitation force $F$ by
\begin{equation}
    u = \frac{F}{m''S} \sum_{n}^\infty  \frac{\psi_n^2(x,y)}{\Lambda_n (\omega_n^2(1+\mathrm{i}/Q_s) - \omega^2)}
    \ ,
    \label{eq:PTL_response_point}
\end{equation}
where $S$ is the area of the substrate and $m'' = \rho_sh_s$ is the mass per unit area of the substrate with $\rho_s$ and $h_s$ being the substrate density and thickness, respectively \cite{Cremer2005}. $\psi_n(x,y)$ are the eigen-modes of the substrate with eigen-frequencies $\omega_n$ and $\Lambda_n = \iint \psi_n^2(x,y)\ \mathrm{d}x\mathrm{d}y/S$. To simplify the model an effective spatial overlap is assumed, $\psi_n^2(x,y) \to \Lambda_n$. Furthermore, a new variable is introduced describing the spectral distance $\Delta\omega_n$ between the excitation frequency and substrate modes defined as $\omega_n = \omega + \Delta\omega$. When only the closest substrate is considered, an effective spring constant describing the coupling between a resonator and the substrate can be derived as
\begin{equation}
	k_\textrm{PTL} = m''S\omega\left(2\Delta\omega + \mathrm{i}\frac{\omega}{Q_s}\right)
	\ .
	\label{eq:k_PTL_domega}
\end{equation}
When a large substrate is used (like a 100 mm silicon wafer) the spectral distance to the closest mode will be difficult to estimate, due to the high density of modes. Instead, a stochastic approach is used where  
$\Delta\omega \equiv \frac{XY}{2}$. Here, $X$ is a uniformly distributed variable between 0 and 1, and $Y$ is an exponential distribution with the mean $n_s^{-1} = \left(\frac{S}{\pi}\sqrt{\frac{3\rho_s(1-\nu_s^2)}{4E_sh_s^2}}\right)^{-1}$, where $n_s$ is the modal density of the substrate. $E_s$ and $\nu_s$ are the substrate's Young's modulus and Poisson's ratio. This leads to a distribution of possible stiffness for the spring.

Due to the assumption that the wavelength of the excited wave in the substrate is much larger than the resonator dimensions, we can treat the outer boundary of the resonator as a single rigid frame with only 1 degree of freedom normal to the plane of the resonator. This frame is then coupled to a reference via a spring defined by $k_\textrm{PTL}$. This enables easy implementation of the PTL model into finite element models. For simulations the expectation value of $k_\textrm{PTL}$ is used (see Eq \eqref{eq:k_PTL}). 
Note that this implementation is only valid for the fundamental mode (and possibly a few higher-order modes) of the resonator. 

\subsection{Stress-thickness dependency}
\begin{figure}
    \centering
    \includegraphics{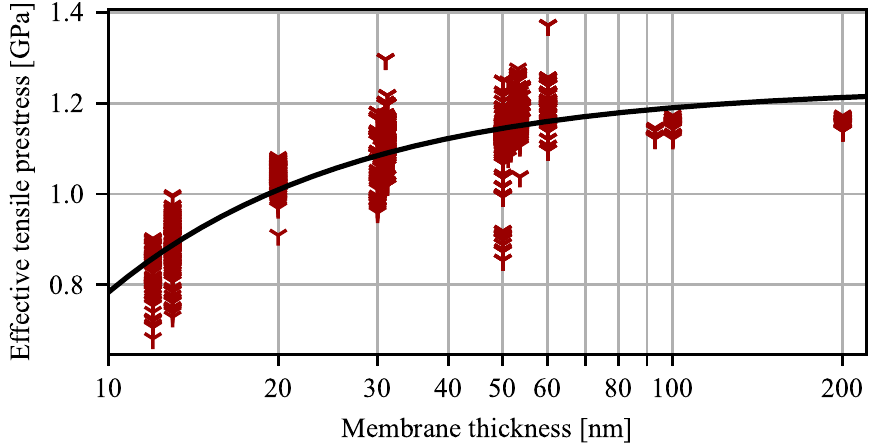}
    \caption{Extraction of the stress-thickness relation of stoichiometric silicon nitride based on frequency measurements of 2573 samples.}
    \label{fig:sigma_fit}
\end{figure}


To model the frequency dependency of the silicon nitride thickness in Fig. \ref{fig:TOT_exp}a, the tensile pre-stress dependency of the thickness is needed. The stress-thickness dependency is believed to be caused by the oxidization layer which introduces a compressive stress contribution onto the silicon nitride film dependent on its thickness \cite{Raider1976}. Assuming that the oxidized layer is much smaller than the total film thickness, we model the effect by the expression $\sigma(h) = \sigma_0 - \beta_\sigma/h$ where $\sigma_0$ is the asymptotic pre-stress parameter and $\beta_\sigma$ is a coefficient that determines how fast the pre-stress changes with thickness. We fit these two parameters against data attained from the measurement of tensile pre-stress from 2573 samples of different thicknesses as shown in Fig. \ref{fig:sigma_fit}. The tensile stress was derived by measuring the resonance frequency and comparing to predicted values from finite element simulations noting the $f\propto\sqrt{\sigma}$ dependency. This approach has some inherent uncertainties related to fabrication and the assumptions of the material parameters of silicon nitride. We find $\sigma_0 = 1.235\pm0.002$ GPa and $\beta_\sigma = 4.52 \pm 0.06 \ \mathrm{Pa\cdot m}$.

\end{document}